\documentstyle[12pt]{article}

\catcode`\@=11
\def\section{\@startsection {section}{1}{\z@}{3.ex plus 1ex minus
 .2ex}{2.ex plus .2ex}{\raggedright\large\bf}}
\def\subsection{\@startsection{subsection}{2}{\z@}{2.75ex plus 1ex minus
 .2ex}{1.5ex plus .2ex}{\raggedright\bf}}
\def\appendix{{\newpage\section*{Appendices}}\let\appendix\section%
        {\setcounter{section}{0}
        \gdef\thesection{\Alph{section}}}\section}
\catcode`\@=12
\newskip\humongous \humongous=0pt plus 1000pt minus 1000pt

\newif\ifdtup

\def\oldreffmt#1{\rlap{[#1]} \hbox to 2\parindent{}}

\def\figfmt#1{\rlap{Figure {#1}} \hbox to 1in{}}

\def\gtap{\raisebox{-.4ex}{\rlap{$\sim$}} \raisebox{.4ex}{$>$}}

\def\beq{\begin{equation}}
\def\eeq{\end{equation}}
\def\bea{\begin{eqnarray}}

\def\com#1#2{
        \left[#1, #2\right]}
\def\eea{\end{eqnarray}}
\def\ap#1,#2,#3#4{           {\it Ann. Phys. (NY)\/ }{\bf #1} (19#3#4) #2}
\def\apj#1,#2,#3#4{          {\it Astrophys. J.\/ }{\bf #1} (19#3#4) #2}
\def\apjl#1,#2,#3#4{         {\it Astrophys. J. Lett.\/ }{\bf #1} (19#3#4) #2}
\def\app#1,#2,#3#4{          {\it Acta Phys. Polon.\/ }{\bf #1} (19#3#4) #2}
\def\com#1,#2,#3#4{          {\it Comm. Math. Phys.\/ }{\bf #1} (19#3#4) #2}
\def\ib#1,#2,#3#4{           {\it ibid.\/ }{\bf #1} (19#3#4) #2}
\def\nat#1,#2,#3#4{          {\it Nature (London)\/ }{\bf #1} (19#3#4) #2}
\def\np#1,#2,#3#4{           {\it Nucl. Phys.\/ }{\bf B#1} (19#3#4) #2}
\def\npps#1,#2,#3#4{         {\it Nucl. Phys. B (Proc. Suppl.)\/ }{\bf B#1}
                             (19#3#4) #2}
\def\pl#1,#2,#3#4{           {\it Phys. Lett.\/ }{\bf #1B} (19#3#4) #2}
\def\pla#1,#2,#3#4{          {\it Phys. Lett.\/ }{\bf #1A} (19#3#4) #2}
\def\pr#1,#2,#3#4{           {\it Phys. Rev.\/ }{\bf D#1} (19#3#4) #2}
\def\prep#1,#2,#3#4{         {\it Phys. Rep.\/ }{\bf #1} (19#3#4) #2}
\def\prl#1,#2,#3#4{          {\it Phys. Rev. Lett.\/ }{\bf #1} (19#3#4) #2}
\def\pro#1,#2,#3#4{          {\it Prog. Theor. Phys.\/ }{\bf #1} (19#3#4) #2}
\def\rmp#1,#2,#3#4{          {\it Rev. Mod. Phys.\/ }{\bf #1} (19#3#4) #2}
\def\sp#1,#2,#3#4{           {\it Sov. Phys.-Usp.\/ }{\bf #1} (19#3#4) #2}

\def\zp#1,#2,#3#4{           {\it Zeit. fur Physik\/ }{\bf #1} (19#3#4) #2}
\catcode`\@=11
\def\@citex[#1]#2{%
\if@filesw \immediate \write \@auxout {\string \citation {#2}}\fi
\@tempcntb\m@ne \let\@h@ld\relax \def\@citea{}%
\@cite{%
  \@for \@citeb:=#2\do {%
    \@ifundefined {b@\@citeb}%
      {\@h@ld\@citea\@tempcntb\m@ne{\bf ?}%
      \@warning {Citation `\@citeb ' on page \thepage \space undefined}}%
      {\@tempcnta\@tempcntb \advance\@tempcnta\@ne%
      \@tempcntb\number\csname b@\@citeb \endcsname \relax%
      \ifnum\@tempcnta=\@tempcntb 
        \ifx\@h@ld\relax%
          \edef \@h@ld{\@citea\csname b@\@citeb\endcsname}%
        \else%
          \edef\@h@ld{\ifmmode{-}\else--\fi\csname b@\@citeb\endcsname}%
        \fi%
      \else
        \@h@ld\@citea\csname b@\@citeb \endcsname%
        \let\@h@ld\relax%
      \fi}%
    \def\@citea{,\penalty\@highpenalty\,}%
  }\@h@ld
}{#1}}

\def\@citeb#1#2{{[#1]\if@tempswa , #2\fi}}
\def\@citeu#1#2{{$^{#1}$\if@tempswa , #2\fi }}
\def\@citep#1#2{{#1\if@tempswa , #2\fi}}

\def\bcites{         
	\catcode`\@=11
	\let\@cite=\@citeb
	\catcode`\@=12
}

\def\upcites{         
	\catcode`\@=11
	\let\@cite=\@citeu
	\catcode`\@=12
}

\def\plaincites{      
	\catcode`\@=11
	\let\@cite=\@citep
	\catcode`\@=12
}

\let\@cite=\@citeb 		

\def\eqnarray{\stepcounter{equation}\let\@currentlabel=\theequation
\global\@eqnswtrue
\global\@eqcnt\z@\tabskip\@centering\let\\=\@eqncr
\gdef\@@fix{}\def\eqno##1{\gdef\@@fix{##1}}%
$$\halign to \displaywidth\bgroup\@eqnsel\hskip\@centering
  $\displaystyle\tabskip\z@{##}$&\global\@eqcnt\@ne
  \hskip 2\arraycolsep \hfil${##}$\hfil
  &\global\@eqcnt\tw@ \hskip 2\arraycolsep $\displaystyle\tabskip\z@{##}$\hfil
   \tabskip\@centering&\llap{##}\tabskip\z@\cr}

\def\@@eqncr{\let\@tempa\relax
    \ifcase\@eqcnt \def\@tempa{& & &}\or \def\@tempa{& &}
      \else \def\@tempa{&}\fi
     \@tempa \if@eqnsw\@eqnnum\stepcounter{equation}\else\@@fix\gdef\@@fix{}\fi
     \global\@eqnswtrue\global\@eqcnt\z@\cr}

\newcount\hour
\newcount\minute
\newtoks\amorpm
\hour=\time\divide\hour by 60
\minute=\time{\multiply\hour by 60 \global\advance\minute by-\hour}
\edef\standardtime{{\ifnum\hour<12 \global\amorpm={am}%
	\else\global\amorpm={pm}\advance\hour by-12 \fi
	\ifnum\hour=0 \hour=12 \fi
	\number\hour:\ifnum\minute<10 0\fi\number\minute\the\amorpm}}
\edef\militarytime{\number\hour:\ifnum\minute<10 0\fi\number\minute}

\def\draftlabel#1{{\@bsphack\if@filesw {\let\thepage\relax
   \xdef\@gtempa{\write\@auxout{\string
      \newlabel{#1}{{\@currentlabel}{\thepage}}}}}\@gtempa
   \if@nobreak \ifvmode\nobreak\fi\fi\fi\@esphack}
        \gdef\@eqnlabel{#1}}
\def\@eqnlabel{}
\def\@vacuum{}
\def\marginnote#1{}
\def\draftmarginnote#1{\marginpar{\raggedright\scriptsize\tt#1}}
\def\draft{
	\pagestyle{plain}
	\overfullrule=2pt
        \oddsidemargin -.5truein
        \def\@oddhead{\sl \phantom{\today\quad\militarytime} \hfil
        \smash{\Large\sl DRAFT} \hfil \today\quad\militarytime}
        \let\@evenhead\@oddhead
        \let\label=\draftlabel
        \let\marginnote=\draftmarginnote
        \def\ps@empty{\let\@mkboth\@gobbletwo
        \def\@oddfoot{\hfil \smash{\Large\sl DRAFT} \hfil}
        \let\@evenfoot\@oddhead}
        \def\@eqnnum{(\theequation)\rlap{\kern\marginparsep\tt\@eqnlabel}%
        \global\let\@eqnlabel\@vacuum}  }

\def\theequation{\thesection.\arabic{equation}}
\@addtoreset{equation}{section}
\@addtoreset{footnote}{section}

\catcode`\@=12

\input epsf
\def\cpsbox{\epsfcheck\cpsbox}
\def\epsfcheck{\ifx\epsfbox\UnDeFiNeD
	\message{(NO epsf.tex, FIGURES WILL BE IGNORED)}
	\gdef\cpsbox##1##2{\vbox to 2in{\hbox to ##1 {\hss} \vss}}
\else\gdef\cpsbox##1##2{
	\setlength{\epsfxsize}{##1}
	\centerline{\epsfbox{##2}}}\fi}

\def\lae{\smash{\,\lower .5 ex \hbox{$\,\stackrel<\sim\,$}}}
\def\gae{\smash{\,\lower .5 ex \hbox{$\,\stackrel>\sim\,$}}}

\def\beq{\begin{equation}}
\def\eeq{\end{equation}}

\begin{document}
\begin{titlepage}
\begin{center}
\today\hfill    WIS--94/27/Jul--PH\\
\hfill       hep-ph/9407324

\vskip 1 cm

{\large \bf  Naturally Light Leptoquarks}

\vskip 1 cm

Ernest Baver and Miriam Leurer\footnote{Address after October 1: Elta, P.O. Box
330, Ashdod, Israel}

\vskip 1 cm

{\em Department of Particle Physics\\
The Weizmann Institute\\
Rehovot 76100\\
ISRAEL}

\end{center}

\vskip 1 cm

\begin{abstract}
Light first generation leptoquarks are being hunted for in HERA and at FNAL
and there are various proposals for further searches in future machines. Such
leptoquarks are however problematic from a theoretical point of view: Low
energy
precision measurements imply strong constraints on the couplings of
the leptoquarks, and up till now the fulfilment of  these constraints seemed
extremely unnatural. Here we show that  horizontal
symmetries, which are very conventional and widely used in the
literature for  completely different purposes, can suppress the
unwanted couplings.
Therefore light first generation leptoquarks
can be natural.
\end{abstract}

\end{titlepage}
\newpage
\section{Introduction}

There has been an increasing interest in light first generation leptoquarks in
recent years, due
to the exciting possibility of observing such particles in the electron-proton
machine HERA \cite{HERA}.
The search for low-lying leptoquarks is supported theoretically by
many beyond-standard models which predict their existence \cite{beyond}.
However, close
phenomenological studies show that leptoquarks are troublesome
\cite{PS,GG,Shanker,Buch,Leurer1,Leurer2,Bendavid}: They can induce
proton decay and  various FCNC processes, and they can enhance leptonic decays
of pseudoscalar mesons by many orders of magnitude. Up to now, these potential
problems with leptoquarks have been
circumvented simply by demanding that the leptoquarks couplings obey a list of
strong constraints.  Here we  point out symmetries
which {\it naturally} lead to the fulfilment of all the desired constraints.
The symmetries we propose are not invented for the sake of leptoquarks but are
rather
conventional symmetries which have been used previously in the
literature for other purposes.

The paper is organized as follows: In the next section we describe in some
detail the phenomenological troubles that leptoquarks may lead to and the
consequent list of constraints on their couplings. We also discuss the general
properties of the symmetries that we propose for the suppression of the
unwanted
couplings. Section 3 contains a specific model that incorporates such
symmetries. Section 4 lists the leptoquarks couplings in the model and section
5 discusses the success of the model in evading the
 phenomenological constraints.
The last section is a short summary and overview.

\section{The unwanted leptoquark couplings}

The processes that lead to the strongest bounds on the leptoquark mass ($M$)
and couplings ($g$) are:
\newline
(i) Proton decay \cite{GG} which is induced when the leptoquark has
also diquark couplings.  The proton decay bound,  $M/g\gtap 10^{16}~$GeV,
can be  avoided by requiring that the  diquark couplings vanish.
\newline
(ii) Flavour changing neutral current (FCNC) processes, which are induced
when the leptoquark couples to a few generations in either the lepton or the
quark sector. The strongest bound arises from $K_L\longrightarrow\mu e$
\cite{PS,Bendavid} and is
typically  $M/g\gtap 100~$TeV. The FCNC  bounds are circumvented
by demanding that the leptoquarks couple ``diagonally'', namely,
they couple to a single generation in the lepton sector and to a single
generation in the quark sector. The diagonality requirement cannot always be
fully satisfied \cite{Leurer1,Leurer2}: In the case of leptoquarks that couple
 to left-handed quarks,
the CKM rotation always induces some nondiagonality in the quark sector,
leading
to an {\it unavoidable} FCNC bound  $M/g^2 \gtap$ few TeV.
\newline
(iii) Enhancement of the leptonic decays of the pion and other pseudoscalars.
The bound is particularly strong for leptoquarks that couple to
both left-handed (LH) and right-handed (RH) quarks
\cite{Shanker,Buch,Leurer2}:
$M/g\gtap 100~$TeV. It is avoided   by demanding that the
leptoquarks couple chirally, namely that they couple {\it either} to LH
{\it or} to RH
quarks  but not to both. Even when the chirality requirement is satisfied,
there are some leptoquarks (those that couple to LH quarks and to LH leptons)
which still contribute significantly to leptonic $\pi$ decay
\cite{Buch,Leurer2}. This
contribution leads to an {\it unavoidable} bound: $M/g\gtap$ few TeV.
\newline
(iv) Atomic parity violation: Any first generation leptoquark induces a
significant and {\it unavoidable} new contribution to atomic parity violation
\cite{Leurer2},
and this leads to
bounds which typically read: $M/g\gtap$ few hundred GeV.

Summarizing the above list,  there are some bounds on leptoquarks that are
unavoidable and
cannot be circumvented, all of them at the TeV scale.  But the most severe
bounds, which send the leptoquark scale to $100~$TeV or even to $10^{16}~$GeV
can be avoided if the leptoquarks obey the
following constraints: They do not couple to diquarks and they couple
diagonally and chirally.

What symmetries could suppress all the unwanted couplings? Clearly,
the diquark couplings can be avoided by the conservation of baryon
or lepton number or some combination of these. It is also easy to
protect the nondiagonal couplings in the lepton sector by imposing separate
conservation of the three lepton numbers: electron, muon and tau. The message
of
our work is that the nondiagonal couplings in the quark sector as well as the
nonchiral couplings can be suppressed by the horizontal symmetries which are
widely discussed in the literature as an explanation for the pattern and
hierarchy in the fermion mass matrices.

The horizontal symmetries have two important characteristics: (i) They, by
definition, distinguish the generations, and therefore are likely to favour the
leptoquark coupling to a particular generation and suppress its coupling to
other generations. This may force the leptoquark couplings to be ``diagonal''.
(ii) They   typically also distinguish the LH and RH
components of each fermion, namely, the two components carry different
horizontal quantum numbers. Therefore such symmetries favour the coupling of a
particular chiral component of a given quark and suppress the coupling to the
other chirality, so that the leptoquark couplings are forced to be ``chiral''.

In the next section we present a leptoquark model which is supplemented by a
horizontal symmetry that suppresses the unwanted couplings.

\section{A model}

There are five possible scalar leptoquark multiplets, and we chose to
concentrate on the multiplet that is motivated by $E_6$ superstring models.
This leptoquark is a singlet of $SU(2)_W$ and carries $\frac 13$ unit of
electromagnetic charge.

As for our choice of the horizontal symmetry: in addition to its traditional
task of ``explaining'' the pattern and the hierarchy in the fermion mass
matrices, the symmetry should also provide the mechanism for suppressing
unwanted leptoquark couplings. In a  recent series of papers
\cite{LNS1,QSA,LNS2} it was shown that abelian horizontal symmetries combined
with supersymmetry provide a particularly efficient mechanism for suppressing
unwanted couplings.  Furthermore, since we worry about FCNC processes which may
be induced by the leptoquark, we turn to a class of models which were
especially constructed to suppress FCNC, the ``quark-squark-alignment'' (QSA)
models \cite{QSA}.

The model we present here has the following symmetries: The standard
$SU(3)_C\times SU(2)_W\times U(1)_Y$ gauge symmetry, supersymmetry,
baryon number, and a
horizontal symmetry which commutes with SUSY and  includes the three
leptonic quantum numbers (electron, muon and tau) and a $Z_8\times Z_7\subset
U(1)_{H_1}\times U(1)_{H_2}$.

The spectrum includes three families of quarks and leptons, two Higgs
multiplets and two leptoquark multiplets\footnote{If we had had only
one leptoquark multiplet, then the sleptoquark would have made the gauge
symmetry anomalous.}. Table 1 lists the chiral multiplets of our model and
their gauge representations and table 2 lists their horizontal charges
($H_1$, $H_2$).
\begin{table}
\begin{center}
\begin{tabular}{|l|l|c|}\hline
Supermultiplet&particle content&gauge representation\\ \hline
$Q_i$&($\tilde q_i$, $q_i$) & $(3,2)_{+{1\over 6}}$\\ \hline
$L_i$&($\tilde l_i$, $l_i$) & $(1,2)_{-{1\over 2}}$\\ \hline
$U^c_i$&($\tilde u^c_i$, $u^c_i$) & $(\bar 3,1)_{-{2\over 3}}$\\ \hline
$D^c_i$&($\tilde d^c_i$, $d^c_i$) & $(\bar 3,1)_{+{1\over 3}}$\\ \hline
$E^c_i$&($\tilde e^c_i$, $e^c_i$) & $(1,1)_{+1}$\\ \hline
$\Phi_u$&($\phi_u$, $\tilde \phi_u$) & $(1,2)_{+{1\over 2}}$\\ \hline
$\Phi_d$&($\phi_d$, $\tilde \phi_d$) & $(1,2)_{-{1\over 2}}$\\ \hline
$S$&($s$, $\tilde s$) & $(\bar 3,1)_{+{1\over 3}}$\\ \hline
$S'$&($s'$, $\tilde s'$) & $(3,1)_{-{1\over 3}}$\\ \hline
\end{tabular}
\medskip
\caption[table1]{
\it The chiral spectrum of the model.
Supersymmetric multiplets (in the first column) are denoted by
capital letters. ``Particles'' and ``sparticles'' (in the second column) are
denoted by the corresponding small letters, and the sparticles are further
provided with a tilde. Here particles are the standard model fermions, the
Higgs scalars and the leptoquark. The subscript $i=1,2,3$ is a generation
index.}
\end{center}
\end{table}

\begin{table}
\begin{center}
\begin{tabular}{lll}
$Q_1(3,0)$ & $Q_2(0,1)$ & $Q_3(0,0)$ \\
$L_1(1,1)$ & $L_2(3,0)$ & $L_3(3,3)$ \\
$U^c_1(-2,3)$ & $U^c_2(1,0)$ & $U^c_3(0,0)$ \\
$D^c_1(-1,2)$ & $D^c_2(4,-1)$ & $U^c_3(0,1)$ \\
$E^c_1(1,2)$ & $E^c_2(2,0)$ & $E^c_3(0,4)$ \\
$\Phi_u(0,0)$ & $\Phi_d(0,0)$ & \\
$S(4,-1)$ & $S'(-4,-6)$ &\\
\end{tabular}
\medskip
\caption[table2]{
\it The ($H_1$, $H_2$) quantum numbers of the matter multiplets in the model.
The $S$ and $S'$ leptoquark multiplets carry $(-1)$ and $(+1)$ units of
electronic lepton number and no muon or tau number.
The horizontal symmetry commutes with SUSY (it is not an R symmetry). Note that
$U(1)_{H_1}\times U(1)_{H_2}$ is free of QCD anomalies.}
\end{center}
\end{table}

Supersymmetry as well as the horizontal symmetry must be broken at low
energies.
We assume that SUSY is softly broken. As for the horizontal symmetry we follow
\cite{FN,LNS1,QSA,LNS2}
and assume that it is explicitly broken in a perturbative manner namely,
terms that break the horizontal symmetry are allowed but suppressed according
to
the following rule: For any term in the Lagrangian that is carrying nontrivial
horizontal charges, let $m_i$ ($i=1,2$) be its $i$'th horizontal charge
modulo $N_i$\footnote{Since the horizontal symmetries of the Lagrangian
are the discrete $Z_{N_i}$ rather than the full $U(1)_{H_i}$, the symmetry
breaking is quantified by  the $m_i$'s.}.
The term is then suppressed by $\epsilon_1^{m_1}\epsilon_2^{m_2}$. The
$\epsilon_i$'s in our model are:
\beq
\epsilon_1=\lambda,~~~~~~\epsilon_2=\lambda^2\;,
\eeq
where $\lambda = 0.2\approx\sin\theta_c$.

In the quark sector, the model is almost identical to the QSA model
of ref \cite{LNS2}. The main difference is that the up quark horizontal numbers
are
somewhat different, so that the ratio of the up quark and top quark masses fits
better the new results from CDF \cite{CDF}. Our model has all the good features
of the QSA
model of \cite{LNS2}. In particular, the alignment of the quark-squark mass
matrices
ensures that no FCNC troubles arise from the contributions of squark-gluino
loop to $K-\bar K$ and $D-\bar D$ mixing, even if the squarks are not
degenerate.

In the lepton sector, our model provides the leptons with masses of the correct
order of magnitude. The separate conservation of the three lepton number
provides a safe protection against FCNC processes in this sector.

\section{The leptoquark couplings}
In this section we  discuss all the renormalizable couplings of the model
that involve the leptoquarks, both in the superpotential and in the soft
supersymmetry breaking terms.
We  will also see how SUSY together with the horizontal symmetry
suppress the unwanted
couplings.

There are two types of  terms in the supersymmetric potential that involve the
leptoquarks: Bilinear terms, which are particularly important since they
provide mass for the sleptoquarks, and trilinear terms which are responsible
for
the leptoquarks' Yukawa couplings to leptons and quarks.

The bilinear term in the supersymmetric potential is
\beq
MSS'
\label{bilin}\; ,
\eeq
where $M$ is a mass parameter
(which is not suppressed by the horizontal symmetry).
When considering the value of $M$ we must take
into account two opposing requirements: On the
one hand $M$ should not be too small,
since it provides the only contribution to the sleptoquark masses.  These
should not be too light to avoid  conflict with  LEP precision measurements
 \cite{LEP}.
On the other hand, $M$
should not be too large since it contributes to the leptoquark masses and we
are interested in the case of light leptoquarks. We satisfy both requirements
by choosing $M$ to be of the order of a few hundred GeV.

The trilinear terms in the supersymmetric potential are given by:
\beq
{G_L}_{ij}L_iQ_jS ~~ + ~~ {G_R}_{ij}E^c_iU^c_jS'
\label{trilin}
\eeq
The order of magnitude values of the $G_L$ and $G_R$ matrices can be deduced
from the horizontal symmetries of the model. In the quark interaction basis we
find:
\beq
G_L=g\left( \begin{array}{ccc}
	1 & \lambda^7 & \lambda^5 \\
	0 & 0 & 0 \\
	0 & 0 & 0 \\
	   \end{array}\right) \;\;{\rm and}\;\;
G_R=g\left( \begin{array}{ccc}
	\lambda^{15} & \lambda^{12} & \lambda^{11}\\
	0 & 0 & 0 \\
	0 & 0 & 0 \\
	   \end{array}\right) \;,
\label{GLGR}
\eeq
where $g$ is  some typical unsuppressed coupling. The matrices
above do not give the exact values of the matrix elements, only their relative
order of magnitude in powers of $\lambda$. Note that the last two rows of $G_L$
and $G_R$ vanish as a consequence of the separate conservation of the
three lepton numbers.

The $G_R$ couplings are strongly suppressed and can practically be ignored.
This is very useful because soft supersymmetry breaking terms can lead to a
{\it
significant mixture} of the $s$ and ${s'}^*$ leptoquarks. If $G_R$ was not
suppressed we would have found ourselves with very non-chiral leptoquarks
leading to severe problems in leptonic $\pi$ decays.

The  nondiagonal couplings in $G_L$ are so suppressed that they too
can be ignored. However, $G_L$
should still be rotated to the mass basis in both the up and down
quark sectors (see \cite{LNS2}) and in this process new nondiagonal couplings
arise:
\beq
G_L^u=g\left( \begin{array}{ccc}
	1 & \lambda & \lambda^3 \\
	0 & 0 & 0 \\
	0 & 0 & 0 \\
	   \end{array}\right)\;,\;\;\;\;
G_L^d=g\left( \begin{array}{ccc}
	1 & \lambda^5 & \lambda^3 \\
	0 & 0 & 0 \\
   	0 & 0 & 0 \\
	   \end{array}\right)\;.
\label{Gud}
\eeq
Note that even after the rotation to the mass basis the nondiagonal terms in
the
down sector are very suppressed and have no practical significance. This is a
result of our using a QSA model, where the rotation between the interaction
basis and the down mass basis is particularly small. This is useful because
FCNC bounds are especially strict for the down-like quarks.

The only couplings of phenomenological importance in (\ref{Gud}) are
the diagonal
${G_L^u}_{11}$ and ${G_L^d}_{11}$ which contribute to $\pi\longrightarrow e\nu$
decay and atomic parity violation, and the nondiagonal ${G_L^u}_{12}$ which
contributes to
$D-\bar D$ mixing.  Identifying $g\equiv {G_L}_{11}$, we find, up to
corrections of order $\lambda^2$:
\begin{eqnarray}
|{G_L^u}_{11}|=|{G_L^d}_{11}|=|g|\nonumber\\
|{G_L^u}_{12}|=|g|\sin\theta_C\;,\label{couplings}
\end{eqnarray}
where $\theta_C$ is the Cabibbo angle.

Turning to the supersymmetry breaking terms: There are three point vertices
involving a squark, a slepton and a leptoquark. These do not cause any
phenomenological problems and will not be further discussed. There are also new
contributions to the leptoquarks masses, which we parameterize by:
\beq
(m^2ss' + h.c.) + m_1^2ss^* + m_2^2s'{s'}^* \; ,
\label{addm}
\eeq
where all the parameters $m$, $m_1$ and $m_2$ are of the order of the SUSY
breaking scale (weak scale). The $m$ parameter is
responsible for the mixture of $s$ and ${s'}^*$ leptoquarks but since the
Yukawa couplings of $s'$ are very strongly suppressed this mixing is not
significant and will be ignored { for the sake of
simplicity}. Altogether the $s'$ leptoquark can be ignored from now on:
Its direct couplings to quarks and leptons are negligible and we
ignore its mixing with $s$.

In closing this section, we mention that the Lagrangian of our model has the
standard model gauge symmetry, supersymmetry, the global horizontal symmetry
discussed in the previous section (including the three separate lepton numbers
and baryon number), and {\it no other continuous accidental symmetry}. We
therefore do not need to worry about the possibility of (pseudo) Goldstone
bosons in the low lying spectrum.

\section{``Measuring'' the success of the model}
In this section we will ``test'' our model, namely we will check to what
extent the bounds on the leptoquark parameters are evaded.
First, we recall that there are  TeV scale bounds that {\it cannot be avoided}
so that the main task
of the symmetries in the model was to get rid  of the
higher scale bounds (Pati-Salam scale, GUT scale). It is easy to see that
here our model is indeed successful: The GUT scale bounds are
avoided  because diquark couplings do not exist. The Pati-Salam
scale bounds are also avoided: the bounds from leptonic $\pi$ decays is
circumvented because nonchiral couplings are very severely suppressed. The
bounds from FCNC processes are also avoided because the couplings in the lepton
sector
are exactly diagonal, in the down quark sector the deviation from diagonality
is so small that it can safely be ignored, and in the up quark sector
nondiagonal couplings are suppressed by $\sin\theta_C$, and consequently
the FCNC bounds from this sector are at the TeV scale (see
\cite{Leurer1,Leurer2}).

Now that we are assured that the bounds on the $s$ leptoquark are at the
``unavoidable scale'', we go into a more detailed test: We will compare the
actual values of the bounds on $s$ in our model with the unavoidable bounds
that
apply to {\it any} leptoquark which is in the $(\bar 3,1)_{\frac
13}$ representation of $SU(3)_C\times SU(2)_W\times U(1)_Y$ and couples to LH
quarks.
The absolutely unavoidable bounds  were studied
in  \cite{Leurer2}:
One arises from leptonic $\pi$ decays (there is also a weaker
bound from atomic parity violation):
\beq
M/g\geq 3.4~{\rm TeV~~~~ at~~ 95\%~CL}\;.
\label{un1}
\eeq
The other bound arises from  FCNC processes which are unavoidable for
leptoquarks that couple to LH quarks. By a fine-tuned division of
the FCNC between the down and up sectors one can minimize the FCNC bound to:
\beq
M/g^2\geq 2.8~{\rm TeV~~~~ at~~ 95\%~CL}\;.
\label{un2}
\eeq

Turning to the corresponding bounds on the $s$ leptoquark in our model we note
that the leptonic $\pi$ decay bound on $s$ is identical to (\ref{un1}).
The FCNC
bound on $s$ arises from $D-\bar D$ mixing. There are two contributions to this
process: One of them arises from a leptoquark-electron box diagram and
 was already
discussed in \cite{Leurer1,Leurer2}. The other contribution
arises from a  sleptoquark-selectron loop. Adding the two contributions
we find:
\beq
\Delta M_D=\frac{1}{192\pi^2}g_L^2\sin\theta_C^2f_D^2M_D^2\frac{1}{M_s^2}(1+
F(x,y))\; ,
\label{DD}
\eeq
where $M_D$ and $f_D$ are the $D^0$ mass and decay constant respectively;
$M_s$ is the
leptoquark mass; $x= (M_{\tilde s}/M_s)^2$ and
$y=(m_{\tilde e}/M_s)^2$ with $M_{\tilde s}$ and $m_{\tilde e}$
being the sleptoquark and selectron masses and
\beq
F(x,y)=\frac{x^2-y^2+2xy\ln{\frac{y}{x}}}{(x-y)^3}
\label{F}
\eeq
parametrizes the sleptoquark-selectron loop contribution.
We wish to translate (\ref{DD}) to a bound on the leptoquark parameters.
To this end we must estimate $F(x,y)$. It is straightforward to show that:
\beq
\frac{2}{3(x+y)}\leq F(x,y)\leq \frac{1}{x+y}\; .
\label{Fbounds}
\eeq
Note that $(x+y)$ is likely to be $\sim 1$ since $\tilde s$ gets its mass
from the supersymmetric parameter $M$, $\tilde e$ gets its mass from soft
supersymmetry breaking terms while $s$ gets its mass from both sources.
It is therefore reasonable to estimate that
\beq
F(x,y)\sim 1\;.
\label{est}
\eeq
Substituting this estimate in $\Delta M_D$ and using the experimental bound
$\Delta M_D < 1.5\cdot 10^{-4}~$eV at $95\%~$CL we find the bound:
\beq
\frac{M_s}{g^2}\geq 6.3~{\rm TeV}\;.
\label{bound1}
\eeq
Comparing (\ref{bound1}) to (\ref{un2}) we see
 that the FCNC bound on our $s$ leptoquark is somewhat
stronger than the unavoidable FCNC bound.  We wish however to stress
that one is not likely to  do better in any model which
is {\it natural} and {\it supersymmetric}. This is due to two
reasons:
\newline
(i) The minimal bound in (\ref{un2}) is achieved by fine tuned
balance between FCNC processes in the down and up sector
\cite{Leurer1,Leurer2}. Since we are
interested in presenting a model which is {\it natural} and not fine-tuned, we
cannot have such a balance. At best, we can ``clean'' one  of the two
quark sectors of FCNC as we did:  By using a  QSA model we
avoided the more severe FCNC of  the down sector, and were left to deal only
with FCNC in the up sector.
\newline
(ii) In a supersymmetric model there is always an additional contribution
to $D-\bar D$ mixing from the sparticle loop, and this necessarily
strengthens the FCNC bound.

Another important remark on the bound (\ref{bound1}) is that it is irrelevant
for light leptoquarks: The $\pi$ decay bound (\ref{un1}) being {\it linear}
in the coupling $g$ rather than quadratic, dominates when the leptoquark mass
is below  $1.9~$TeV.
Therefore, the bound on the leptoquark parameters is just the
minimal unavoidable bound if $M_s<1.9~$TeV.

Summarizing, our model  avoids all the  high scale bounds on the leptoquark
parameters.  For leptoquark masses below $1.9~$TeV the model does not impose
any further bound beyond the absolutely unavoidable (\ref{un1}), while for
heavier leptoquarks the dominant bound is (\ref{bound1}).  This latter bound
is somewhat stronger than the unavoidable (\ref{un1},\ref{un2}), but we think
it unlikely that it could be improved in any natural supersymmetric model.

\section{Summary and outlook}

We showed that the horizontal symmetries which are usually used to explain the
pattern and hierarchy in the quark mass matrices can also be used to suppress
unwanted leptoquark couplings. We presented a particular model of a first
generation leptoquark with a horizontal symmetry and found that the unwanted
couplings are naturally suppressed and the phenomenological constraints that
may arise from the unwanted couplings are almost completely circumvented.

In concluding, we wish to point  to a possible
extension in the use of the
horizontal symmetry protection mechanism.
We showed that the
horizontal symmetries can protect
nonchiral couplings and nondiagonal couplings in the quark sector but they
could actually be used to protect also the nondiagonal couplings in
the lepton sector and  it is
possible that one may even progress one step further,
drop altogether the conservation of all lepton and baryon numbers and provide
sufficient protection to all the unwanted couplings via horizontal symmetries.

{\noindent {\bf Acknowledgements:} We thank Neil Marcus and Yossi Nir for
useful
remarks.}

\end{document}